\documentclass[prl,twocolumn]{revtex4}

\usepackage{mathrsfs,amsmath,graphicx}

\def\scr{\mathscr}

\def\SL{{\scr L}}   
 \def\SD{{\scr D}}  
 \def\SI{{\scr I}}  
   \def\SP{{\scr P}}

\def\Bid{{\mathchoice {\rm {1\mskip-4.5mu l}} {\rm
{1\mskip-4.5mu l}} {\rm {1\mskip-3.8mu l}} {\rm {1\mskip-4.3mu l}}}}

\def\avg#1{\langle#1\rangle}    \def\<{\langle}         \def\>{\rangle} 
            
    \def\tr{{\rm tr}}

\def\al{\alpha}         \def\bt{\beta}          

\def\sig{\sigma}        \def\del{\delta}        \def\Del{\Delta}
\def\eps{\epsilon}        
\def\up{\uparrow}       \def\down{\downarrow}

    \def\VD{\boldsymbol{D}}
  \def\V0{{\mathbf 0}}

\def\Bx{{\mathbf x}}  \def\B0{{\mathbf 0}}

 \def\Bp{{\mathbf p}} \def\Bq{{\mathbf q}}

\def\BJ{{\mathbf J}}

\def\td#1{{\tilde{#1}}}

\def\be{\begin{equation}}       \def\ee{\end{equation}}
\def\bea{\begin{eqnarray}}      \def\eea{\end{eqnarray}}
\def\nn{\nonumber}

\begin{document} 

\title{Effective theory of excitations in a Feshbach resonant superfluid}

\author{W. Vincent Liu}
\affiliation{Department of Physics and Astronomy, University of
  Pittsburgh, Pittsburgh, Pennsylvania 15260}


\begin{abstract}
 
A strongly interacting Fermi gas, such as that of cold atoms operative
near a Feshbach resonance, is difficult to study by perturbative
many-body theory to go beyond mean field approximation. Here I
develop an effective field theory for the resonant superfluid based
on  broken symmetry. The theory retains both fermionic
quasiparticles and superfluid phonons, the interaction between them
being derived non-perturbatively.  The theory converges and can be
improved order by order, in a manner governed by a low energy
expansion rather than by coupling constant.  I apply
the effective theory to calculate the specific heat and propose a
mechanism of understanding the empirical power law of energy versus
temperature recently measured in a heat capacity experiment.

\end{abstract}

\maketitle

In the last few years, the rapid development of experiments of
atomic Fermi gas using Feshbach resonance has revitalized
theoretical interest of strongly interacting many-body system. 
Recent experiments, for instance,
the collective excitation~\cite{Bartenstein+Grimm:04,Thomas_radial:04}
and  heat capacity~\cite{Kinast:05}, point to the need of a
dynamical theory  beyond mean-field BEC-BCS crossover theory.  
A basic difficulty with the
crossover theory is that
perturbative expansion is not justifiable if
going to next order for strong interaction.  
Various efforts have been made on the basis of 
hydrodynamic theory~\cite{Stringari:04,Heiselberg:04,Hu++Tosi:04,Kim-Zubarev:05,Bulgac+:05},  
which in general explain the frequency of collective modes
particularly well
on the resonance point.  A disadvantage
of hydrodynamic theory is that fermionic excitation is hard to
incorporate beyond the density and velocity field approximation.  In
this paper, I develop an effective field theory for the resonant
superfluid that retains both fermionic and bosonic excitations
explicitly based on broken symmetry. An advantage of this theory over
a pure bosonic theory is that it allows one to study, for example, the
damping of collective modes due to scattering with fermionic
quasiparticles.  The present approach is thus different from and
complimentary to the existing crossover and hydrodynamic theories.

\paragraph{Effective field theory of excitations.}
The physical system we are interested in is a ultra-cold gas of
fermionic atoms (such as $^6$Li or $^{40}$K) with two internal (hyperfinespin)
states in a broad Feshbach
resonance. It is known from both theory and experiment
(see, for instance,
\cite{Szymanska++Burnett:05} and \cite{Partridge++Hulet:05},
respectively) 
that the  molecule
fraction is small from the resonance regime through the entire fermionic
side.  
The microscopic theory is then described by the Lagrangian (in the
path-integral framework with imaginary time, $\tau=it$):
\be
L= \int d^3\Bx \left\{\psi^*_\sig \left(\partial_\tau-{\textstyle{\nabla^2\over
  2m}}-\mu\right)\psi_\sig + g
\psi^*_\up\psi^*_\down
\psi_\down\psi_\up\right\} 
\ee
where $\psi_\sig(x)$ describes 
fermionic atom fields at $x=(\Bx,\tau)$ and  spin
state   $\sig=\up,\down$, $g$ is the coupling constant 
defined by $g={4\pi a\over m}$ ($a$ the  $s$-wave
scattering length),   
and $\mu$ is the chemical
potential. (Throughout, summation is implicit over repeated
spin indices; $\hbar\equiv 1$ is set
in units.)

For attractive interaction $g$,  
one can perform an {\it exact} Hubbard-Stratonovich
transformation to introduce the auxiliary pair fields, $\Del(x)$, and
its complex conjugate. Then, the Lagrangian becomes
\bea
L&=& \int d^3\Bx \Big\{\psi^*_\sig\textstyle 
\left(\partial_\tau-{\nabla^2\over
  2m}-\mu\right)\psi_\sig \nn \\
&& \textstyle \hspace{2em} + (\psi^*_\up \psi^*_\down \Del +c.c.) +
{1\over g}|\Del(x)|^2  \Big\}  
\,. \label{eq:LpsiDel}
\eea
Up to this point, no approximation has been made. One may proceed then
to derive the gap and chemical equations, following the BCS-BEC
crossover theory procedure (for instance, Ref.~\cite{Engelbrecht:97} and
references therein). But we are interested in the dynamics of low
energy excitations about the superfluid state.

The atomic gas is in strong coupling
regime, having a scattering length greater than or comparable to the Fermi
wavelength $k_F^{-1}$ ($|k_F a|\gtrsim 1$). 
Conventional perturbation theory does not directly apply, at
least not in a standard, controllable manner. To have a controlled theory
for the description of  quantum dynamics beyond mean field theory,
one may seek for available different approach. 

The broken symmetry of the superfluid state provides such an 
approach---effective field theory~\cite{Weinberg:bk96:ch19+Liu}. 
The atomic
gas conserves the total number of particles. 
This conservation law corresponds to a U(1) symmetry of the
phase transformation. For the above model, the U(1) symmetry
transformation is
\be
\psi_\sig(x) \rightarrow e^{i\al} \psi_\sig(x)\,,\quad
\Del(x) \rightarrow e^{i2\al}\Del(x)\,,
\ee
with  $\al$ an arbitrary, constant phase.

Recall that the superfluid state is a Bose-Einstein condensate of
fermionic atom pairs, which breaks the U(1) symmetry.  The order
parameter acquires a finite value below the superfluid phase
transition temperature, $\avg{\Del(x)}=\Del_0\neq 0$. Fluctuations
about this superfluid ground state are
excitations.

Here I derive an effective theory of the phase
fluctuation, directly based on the broken symmetry. We start by
treating $\Del_0$---the groundstate or thermal mean value of the order
parameter---as an input parameter. It may be taken directly from
experiments; of course it can also be taken from the result of Monte
Carlo, or any mean field crossover theory but that will require
further justification.
We are concerned with phase fluctuations in space and time, so write
the order parameter field
\be
\Del(x) = \Del_0 e^{i2\theta(x)}\,.
\ee
(The factor $2$ is inserted to make it explicit that the pair field
carries two units of mass.)
We have set to ignore the amplitude fluctuation for the purpose of
simplicity; this is justifiable
for it is known gapped and, as can be checked below, 
does not directly coupled to phase except
via fermions. 
Instead of making a Gaussian expansion---a usual practice,
we transform {\it non-perturbatively} 
the fermion fields at each space time point as follows:
\be
\psi_\sig(x) = \td{\psi}_\sig(x) e^{i\theta(x)}\,, 
\quad 
\psi^*_\sig(x) = \td{\psi}^*_\sig(x) e^{-i\theta(x)}\,. \label{eq:psi2}
\ee
The transformation is designed to eliminate the phase fluctuation
$\theta(x)$ dependence from the off-diagonal pairing potential
terms, $(\psi^*_\up \psi^*_\down \Del +c.c.)$. 
As a result, the $\theta(x)$-dependence can only arise  from the
kinematic terms of 
the fermion theory sector.   Collecting all terms of
$\td{\psi}$ and $\theta$, we have
\be
\SL_{\td{\psi}\theta} = 
  \td{\psi}^*_\sig (D_\tau
-\mu - \textstyle{{\VD^2\over 2m}} )\td{\psi}_\sig 
+ (\Del_0 \td{\psi}_\down \td{\psi}_\up + c.c.) \label{eq:Lpsitheta}
\ee
where 
$
D_\nu\equiv\partial_\nu + i\partial_\nu\theta(x)\,,\ \nu=\tau, x, y, z\,. 
$
The U(1) symmetry guarantees {\it the phase field to appear only through
derivatives}, since energy cannot depend on a constant phase.
The form of coupling
between the phase and  fermion fields is exact from
the model. 
This theory is no perturbation, apart from excluding the amplitude
fluctuation. 

For those who are familiar
with gauge theory, the time and spatial derivatives of $\theta(x)$
together behave like a four component gauge field.   
After the transformation (\ref{eq:psi2}), the fermions
$\td{\psi}_\sig$  see a constant off-diagonal pairing potential
$\Del(x)=\Del_0$ but interact with the (Goldstone) phase field
$\theta(x)$ by a means similar to  `gauge' coupling.    

\paragraph{Landau superfluid hydrodynamics.} 
The effective theory (\ref{eq:Lpsitheta}) 
has a very strong implication for  superfluid
hydrodynamics. To reveal that, let us collect all terms of first order in
$\theta$ from $\SL_{\td{\psi}\theta}$:
\be
+\hat{n}(x) i\partial_\tau \theta(x) + \nabla\theta(x) \cdot
\hat{\BJ}(x) 
\label{eq:ph-phase}
\ee
where 
$\hat{n}(x) =\td{\psi}^*_\sig(x) \td{\psi}_\sig(x)=\psi^*_\sig(x)
  \psi_\sig(x)$ and $\BJ(x) =-{i\over 2m} (\td{\psi}^\dag_\sig \nabla
\td{\psi}_\sig- \nabla\td{\psi}^\dag_\sig \td{\psi}_\sig)$
are the fermion density and current operators. The first term implies
that the fermion {\it total} density and the order parameter phase are
canonically conjugate.  The amplitude fluctuation of the order
parameter is separate. [This can be  checked by shifting 
$\Del_0$ by a quantum fluctuation $\Del'(x)$, 
$\Del_0\rightarrow \Del_0+\Del'(x)$.]

To derive the hydrodynamic theory, we add to $\SL_{\psi\theta}$ the term
\be
\Del{\SL}= -i \lambda(\rho
  -\psi^*_\sig\psi_\sig) 
\ee
and do path integrals over the new fields $\rho(x),\lambda(x)$. This
is allowed because the path-integral over $\lambda(x)$ gives a
$\del$-functional, which sets the equation
\be
\rho(x)=\psi^*_\sig(x)\psi_\sig(x)\,,  \quad \mbox{(the density field)}
\ee
and the path integral over $\rho(x)$ gives a constant. The equation
justifies  $\rho(x)$ as fermion density field. 
The added term
$\Del\SL$ is designed to extract the effective density field. Up to now
the transformation of $\SL_{\psi\theta}$ is exact.  
Performing path integrals over fermion fields produces the effective
action:
\bea
S&=&S_{\rho\theta}+S_\lambda +\ldots\\
S_{\rho\theta}&=& \textstyle \int_x  \rho\Big[i\partial_\tau\theta
 +  {(\nabla\vartheta)^2\over 2m}\Big] \nn \\
&& \textstyle +\int_{xx'}\SP_{ij}(x-x')
\partial_{x_i} \theta(x) 
\partial_{x'_j} \theta(x')\nn \\
S_\lambda&=& \textstyle
\int_x i\rho\lambda - {1\over 2}\int_{xx'}  \SD(x-x')\lambda(x)
  \lambda(x')
\eea
where `\ldots' stands for higher powers of $\theta$ and/or $\lambda$
fields. 
Here, $\SD$ and $\SP$ are the density and current correlation
functions defined in the homogeneous ground state (so the subscript `$0$'): 
\bea
\SD(x-x')=-\avg{\hat{n}(x)\hat{n}(x')}_0 && \mbox{and} 
\label{eq:Dnn}
\\
\SP_{ij}(x-x')=-\avg{\hat{J}_i(x)\hat{J}_j(x')}_0\,, && i,j=x,y,z.
\label{eq:PJJ}
\eea
This is a low energy (derivative) expansion of the collective
fields, rather than a perturbation in some coupling constant. The
derivative nature of $\lambda(x)$ is less obvious, but immediately
becomes manifest when examining the quantum equation of motion, 
$$
{\del S\over \del \rho }=0 \quad \Rightarrow \quad \lambda = - \partial_\tau
\vartheta
+ {i\over 2}  (\nabla\vartheta)^2 \,.
$$
Therefore the power expansion on $\lambda$ is equivalent to a low
energy expansion. The field 
$\lambda$ is auxiliary in nature. We can calculate the effective
couplings of $\theta$ and $\rho$ 
mediated by $\lambda(x)$ perturbatively in low energy limit. The
resulting effective action is the equivalent of Landau superfluid 
hydrodynamics:
\bea
S^\prime_{\rho,\theta} &=& S_{\rho\theta} 
 -{1\over 2}\int_{xx'}  {\SD(x-x')}^{-1}\rho(x) 
  \rho(x')
\eea 
where $S_{\rho\theta}$ is the same as given above.

\paragraph{Effective theory of phase fluctuations.} 
Single fermion excitations are gapped at all energies below $\Del_0$.
The Goldstone boson $\theta(x)$ interacts with fermion through
particle-hole pairs. Therefore, for collective excitations of energy
below $2\Del_0$, we can integrate out all fermionic degrees of  freedom and
have an effective theory for $\theta$ alone. Starting from the theory
(\ref{eq:Lpsitheta}),  we write it into a quadrature of 
fermion fields in the Nambu spinor space, i.e.,
\bea
\int d^3\Bx d\tau \Big(\psi^*_\up(x), \psi_\down(x)\Big) K_{x,x'}
{\psi_\up(x')\choose \psi^*_\down(x')}
\eea
with 
\bea
 K_{x,x'} &=&\textstyle
 [K^0(x)+ K^\theta(x)]\del^3(\Bx-\Bx')\del(\tau-\tau')\,, \\
 K^0(x) &=& \textstyle \Bid \partial_\tau
-\tau_3 \Big({\nabla^2\over 2m} +\mu\Big) +\tau_1\Del_0\,,\\
K^\theta(x) &=& \textstyle
\Bid {(\nabla\theta)^2-\nabla\theta\cdot\nabla \over 2m}
+ \tau_3 i\partial_\tau\theta\,,  
\eea
where $\{\Bid,\tau_1,\tau_2,\tau_3\}$ are the identity and Pauli
matrices for the Nambu spinor space.  Integrating out fermion fields
gives the effective action for the phase mode $S_\theta=\int_x \tr [\ln
K]_{x,x}$. 

After some
straightforward calculation, the effective action of the theory is
found to be (in momentum-frequency space) 
\bea
S_{\theta} &=& \sum_{q} \textstyle 
\Big[ -{\SD(q)\over 2} \omega^2 
+{\SP_\parallel(q)\over 2}  \Bq^2+ {n_0\over 2m}
\Bq^2\Big]\theta(q)^*\theta(q)
 \nn \\
&& \hspace{0.1em}
   + \textstyle O\big({\omega^4\over (2\Del_0)^4},{\Bq^4\over
     (2\Del_0)^4}\big) \, \mbox{(higher powers of $\theta$)} 
\label{eq:Seff}
\eea
where $n_0$ is the total fermion density and 
\be
\begin{array}{l} \displaystyle
{\SD}(q) 
= {1\over \bt V} \sum_{p} 
 \tr\{ \tau_3 G(p) \tau_3 G(p+q)\} \,,\quad
 q\equiv(\Bq,i\omega)\\ \displaystyle
\SP_\parallel(q) = {{1\over \bt V }} \sum_{p} \Bp_\parallel^2
 \tr\{ G(p-{q\over 2}) G(p+{q\over 2})\} \,,\  \Bp_\parallel\parallel \Bq,
\\ \displaystyle
G(p) =  
{ i\omega^\prime  + \tau_3 \eps_\Bp + \tau_1 \Del_0
\over (i\omega^\prime -E_\Bp) 
(i\omega^\prime+E_\Bp)}   \,,\quad  p\equiv(\Bp,i\omega')\,,
\\
\end{array}\nn 
\ee
with 
$\eps_\Bp={\Bp^2\over 2m} -\mu$ and
$
E_\Bp =  \sqrt{\eps_\Bp^2 +\Del_0^2} \,.
$
$E_\Bp$ is the fermionic quasiparticle spectrum. 
We are working in the imaginary time formalism: 
$\omega$ and $\omega'$ are the Matsubara
frequency for bosons and fermions, respectively;
$\bt V$= space-time volume; and  $G(p)$ is 
the fermion
propagator.
For readers who are familiar with  diagrammatic calculation,
$\SD(q)$ and $\SP(q)$ are directly related 
to the density and current correlation functions defined in 
Eqs.~(\ref{eq:Dnn}) and (\ref{eq:PJJ}) by Fourier transformation. 

One can directly evaluate $\SD(q)$ and $\SP(q)$.  
The calculation is tedious but otherwise
elementary. At zero temperature $T=0$, the result is 
\bea
{\SD(q)\choose {\SP}_{\parallel}(q)}  &=& 
\int_\Bp {1\choose \Bp_\parallel^2} \left[ 1-{\eps_+\eps_- \mp \Del^2_0\over
    E_+E_-} \right]\times \nn \\
&&  {E_+ + E_-\over (i\omega -E_+ -E_-)(i\omega +E_+
  +E_-)} 
 \,, \label{eq:D(q)+P(q)}
\eea
where $\int_\Bp\equiv \int {d^3 p\over (2\pi)^3}$,
and the subscript $\pm$ stands for $\eps_\pm\equiv
\eps_{\Bp\pm\Bq/2}$ and likewise for $E_\pm$. (Note the important sign
difference in the `coherence factor' between
$\SD$ and $\SP$.)

Here to an important point regarding the {\it perturbative
expansion}.    The expansion
is sorted by the powers of momentum and frequency of the field
$\theta(\Bq,i\omega)$. Unlike the Gaussian fluctuation theories, the
present approach 
does not require the phase fluctuation field itself small but 
slowly varying in space and time. This approach is also quantitative: The
low energy and momentum scales are set by $\Del_0$ and $\Del_0/v_F$,
respectively; One can improve the theory  order by
order and estimate the next order correction. A more
complete development of this approach will be given in a future
study. The essential point of why this approach works is that the
phase fluctuation (Goldstone) field must appear 
through derivatives {\it dictated} by the
broken U(1) symmetry. 

For those familiar with field theory, I should be
careful about approximation  in the expansion of
(\ref{eq:Seff}). The general form is no approximation, but the
coefficients $\SD$ and $\SP$, however, could not be derived in any
`honest' calculation for strongly interacting gas
if we had included the
amplitude fluctuation too. So the coefficients should be
treated as good as mean field approximation.  Nowadays, a more
typical application of Weinberg effective field theory in particle
physics is to directly construct ingredients of proper symmetry
transformation, write a general form of effective Lagrangian, and
match the unknown coefficients with experiments rather than derive.
We will not discuss that general approach here. 

\paragraph{The $\omega_\Bq \ll 2\Del_0$ limit.}  This corresponds
to the limit of long wavelength, $\Bq\rightarrow 0$.
Expanding the ${\SD}$ and
${\SP}$ of Eq.~(\ref{eq:D(q)+P(q)}) 
in powers of $\Bq^2$, we get the leading order
\bea
{\SD}(q=0) 
&= & -\int {d^3p\over (2\pi)^3} {\Del_0^2\over E_\Bp^3 }= -{\td{\Del}^2
  mk_F\over \pi^2} \SI \\
{\SP}(q=0)&=&0\,,
\eea
where $\td{\Del} \equiv {\Del_0/ \eps_F}$\,, $\td{\mu}\equiv{\mu/\eps_F}$ and 
$\textstyle
\SI= \int_0^\infty dx {x^2\over 
\left[(x^2-\td{\mu})^2+{\td{\Del}}^2\right]^{3/2}}\,.
$
[$\eps_F$ is defined as the Fermi energy at the non-interacting limit.]
The difference between $\SD$ and $\SP$ can be seen from the coherence
factor in their expressions. In the case of ${\SP}(q)$, the factor is
exactly vanishing at $\Bq=0$. 

Then the effective action of the phase mode becomes
\bea
S_{\theta} &=& \textstyle {1\over 2} \sum_{q} \Big[ {\Del_0^2 mk_F\SI \over
    \pi^2\eps_F^2}  \omega^2 
+ {n_0\over m}
\Bq^2\Big]\theta(q)^*\theta(q)
 \nn \\
&& \hspace{1em}
   + \textstyle O\Big({\omega^4\over (2\Del_0)^4},{\Bq^4\over
     (2\Del_0)^4}\Big) \,. 
\label{eq:Seff2}
\eea
The energy spectrum of the collective mode is obtained by analytical
continuation $i\omega \rightarrow \omega_\Bq +i0^+$ and by finding the
zero of the coefficient of the quadratic $\theta$-term. The spectrum
is  
\be\textstyle
\omega_\Bq = v_s |\Bq| + O\left(\frac{|\Bq|^3}{\Del_0^2}\right)
\,, 
\quad v_s={v_F\over \sqrt{3 \td{\Del}^2\SI}}\,, \label{eq:omega_q}
\ee
where $v_s$ is the sound velocity.   

For small $\Del_0$ compared with the Fermi energy $\eps_F$ (BCS
 limit), 
$
\SI\approx {\eps_F\over \Del_0^2}[1 +O(\Del^2_0)]
$. Using the relation $k_F=(3\pi^2 n_0)^{1/3}$ and $v_F=k_F/m$, 
this reproduces the sound velocity $v_s=v_F/\sqrt{3}$---the well-known
result in the BCS limit obtained in the theories of Bogoliubov,
Anderson, and others later on~\cite{Anderson:58,Engelbrecht:97}.

\paragraph{Specific Heat.}
Recently, Kinast {\it et
al}~\cite{Kinast:05} reported a measurement of the heat
capacity of a strongly interacting Fermi gas of $^6$Li atoms, slightly
detuned to the BCS side from the resonance. The heat capacity reveals
a kink at a temperature about $T=0.27 T_F$, arguably an evidence of superfluid
phase transition.  While the crossover theory employed in the paper 
agrees with the data 
in general, it deviates from the data noticeably at low temperature
below the transition, as shown in their figure of energy input versus
temperature.  

\begin{figure}[htbp]
\includegraphics[width=\linewidth,height=0.5\linewidth]{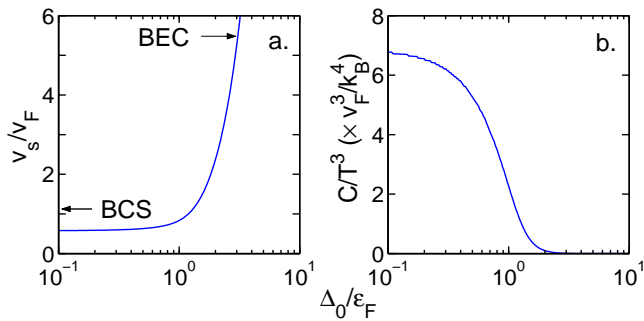}
\caption{Sound velocity and specific heat in a resonant fermionic
  superfluid. The chemical potential is self-consistently determined
  when varying $\Del_0$ with density fixed. Parameters: $k_F=(3\pi^2
  n_0)^{1/3}$; $\eps_F=k_F^2/(2m)$; $v_F=k_F/m$. The BEC limit
  (molecular side) is plotted to just show the trend; otherwise it is
  beyond the scope of validity of the present theory.}
\label{fig:sound_velocity}
\end{figure}

As an example of application, I will use the effective field theory
(\ref{eq:Seff2}) to
calculate the specific heat for temperature
well below the superfluid critical temperature $T_c$ ($T\ll T_c\sim
2\Del_0/k_B$). 
All fermionic quasiparticle
excitations are gapped, so the only contribution to the entropy must
be due to the collective superfluid phonons with a linear dispersion. 
The thermal energy per
unit volume due to the phonons is ${E\over V} ={\pi^2(k_B
T)^4\over 30 v_s^2}$.  The coefficient factor (including $v_s$)
is model dependent; but the $T^4$ power law is universal in the
limit of $T\ll T_c$.  Differentiating the thermal
energy gives the specific heat, $c=\frac{2\pi^2 k_B^4 T^3}{15v_s^3}$.
By measuring the coefficients of the $T^3$ law, one can determine the
sound velocity and check the theoretical result.
Fig.~\ref{fig:sound_velocity} shows how $v_s$ and $c$ vary with
$\Del_0$ between weak and strong coupling limits---a result seeming
not reported anywhere else yet.   

The $T^4$ power law of the energy differs from the empirical $E\propto
T^{3.73}$~\cite{Kinast:05}. Bulgac~\cite{Bulgac:05pre} examined the
role of collective excitations by focusing on the unitary regime and
predicted a complicated $T$-dependence in different temperature
regime, from exponential to $E\propto T^5$.  Here I provide an
explanation other than those discussed in
Ref.~\cite{Kinast:05,Bulgac:05pre}. Away from the limit of $T\ll T_c$,
we now move to the temperature regime slightly below $T_c$ ($0\ll
T\lesssim T_c$). There can be preformed Cooper pairs but the fraction
of condensed Cooper pairs is small. In our effective field theory
description, that corresponds to a small amplitude of order parameter,
$\Del_0$.  The superfluid phonon still exists but looses the linear
dispersion characteristic. This fact is actually implicit in
Eq.~(\ref{eq:omega_q}) for which the expansion ought to include
nonlinear powers of $|\Bq|/\Del_0$ when $\Del_0$ is relatively small.
In fact, for temperature near $T_c$, the expansion should be done
another way around, in terms of small $\Del_0/|\Bq|$ instead. At the
end, the superfluid phonon and non-condensed Cooper pair, coupled
together, all are expected to have a dispersion quadratic in momentum,
$\omega_\Bq\sim \Bq^2$. Such a Bose gas gives a temperature dependence
as $E\propto T^{5/2}$ by simple power counting.  For strongly
interacting gas, fermionic excitations are believed to have a
pseudogap far greater than $\Del_0$---sometimes called coherent
superfluid gap. Therefore fermionic excitations do not alter the above
power law.  (There might be contributions due to gapless fermions on
the trap edge if the effect of trap were included, but that seem not
as important in the thermodynamic limit.)  In summary, the power law
is $T^4$ for $T/T_c\rightarrow 0$ and is argued to be $T^{5/2}$ for
$0\ll T\lesssim T_c$. Perhaps, in my opinion, an interpolation of
these two limits can fit the data better, entirely consistent with the
empirical $T^{3.73}$ law from a single-exponent fit.

All results apply on the fermionic side of a broad Feshbach resonance,
when $\Del_0>0$ (the larger the better, contrary to weak perturbation
theory).  One may easily generalize our approach to develope a more
complete effective theory valid for the molecular side as well, by
starting with a Bose-Fermi resonance model of both atoms and
molecules.

Finally, I learned that the low energy expansion was used long ago to
derive time-dependent Ginzburg-Landau theory for a BCS superconductor
by Abrahams and Tsuneto~\cite{Abrahams:66}. But their expansion was
done in a very different manner and assumed the (weak coupling)
BCS limit which was quite alright at that time. Another useful
difference is that the present effective theory, having retained
fermions explicitly, is convenient to study, for example, the damping
of collective modes due to decay into (fermionic) quasiparticles.

I am grateful for discussions with D. Son, J. Thomas, and F. Zhou. 
I thank the Institute for Nuclear Theory at the University of
Washington for its hospitality through the Program on
Quantum Liquids and Gases and the Department of Energy for partial
support during the completion of this work.


\bibliography{sound_wave}
\end{document}